\documentclass[nofootinbib,preprintnumbers,10pt,amsmath,amssymb]{revtex4}
\usepackage{graphicx}
\usepackage{dcolumn}
\usepackage{bm}
\usepackage{amsfonts}
\usepackage{amsmath}
\usepackage{amssymb}
\usepackage{amsthm}
\usepackage{citesort}

\begin{document}


\title{Notes on Gibbs Measures}

\author{Jinshan Zhang}
\email{zjs02@mails.tsinghua.edu.cn}

\affiliation{
Department of Mathematical Sciences, Tsinghua University, Beijing, China, 100084 
}%

\date{May 10th 2008}

\begin{abstract}
\textbf{Abstract.} These notes are dedicated to whom may be
interested in algorithms, Markov chain, coupling, and graph theory
etc. I present some preliminaries on coupling and explanations of
the important formulas or phrases, which may be helpful for us to
understand D. Weitz's paper ``Combinatorial Criteria for Uniqueness
of Gibbs Measures" with ease.

\
\end{abstract}

\maketitle

\section{Brief Introduction}
The structure of this notes is as follows. Preliminaries on coupling
are proposed in section II. We go on in section III to show some
details of the formulas  and some explanations on the ``remark" in
the paper.

\section{Preliminaries}
In order to show some properties of coupling, above all we discuss
some elementary but important concepts in probability theory. We
only focus our attention on the discrete state space which is enough
for us to read the paper, although all the conceptions and
properties have general definitions and generalizations.\\\\
\textbf{Definition 2.1} (Distance in variation) Let $E$ be a
countable sates-space and let $\mu$ and $\nu$ be two probability
measures on $E$. The distance in variation between $\mu$ and $\nu$
is defined by
\begin{displaymath}
d(\mu,\nu)=\frac{1}{2}\sum\limits_{i\in E}|\mu(i)-\nu(i)|.
\end{displaymath}
Remark 1: If we denote by $\mathcal{M}(E)$ the collection of all the
probability measures on $E$, it's simple to check $d(\mu,\nu)$ is a
true metric on the set $\mathcal{M}(E)$.\\\\
Remark 2: One can easily check $d(\mu,\nu)= \sup\limits_{A\subset
E}|\mu(A)-\nu(A)|= \sup\limits_{A\subset
E}\mu(A)-\nu(A)=1-\sum\limits_{A\subset E}\min(\mu(i),\nu(i))$.\\
(Hints: Let $B=\{i: \mu(i)-\nu(i)\geq 0\}$, then
$\sup\limits_{A\subset E}\mu(A)-\nu(A)=\mu(B)-\nu(B)
=\nu(B^{c})-\mu(B^{c})=\frac{1}{2}\sum\limits_{i\in
E}|\mu(i)-\nu(i)|$.)\\\\
\textbf{Definition 2.2} Let $E$ be a countable state-space, and
$\mu$ and $\nu$ be two probability measure on $E$. Let $X$ and $Y$
be two random variables from the probability space $(\Omega,
\mathcal{F}, P)$ to $(E,\mathcal{B}(E))$, where $\mathcal{B}(E)$
denote the $\sigma$ field generator by all the elements of $E$.
Assume that $X$ and $Y$ have the distribution $\mu$ and $\nu$
respectively, then the joint distribution of the bivariate r.v.
$(X,Y)$ is called a coupling of $\mu$ and $\nu$.\\\\
Remark 1: Any distribution of a bivariate $(X,Y)$ with  marginal
distribution $\mu$ and $\nu$ constructs a coupling of $\mu$ and
$\nu$.\\\\
Remark 2: The exitance of the coupling of $\mu$ and $\nu$ only lies
in the exitance of two r.v. $X$ and $Y$ since we can always
construct a bivariate r.v. $(X,Y)$ with marginal distribution $\mu$
and $\nu$ (e.g. $X$ and $Y$ are independent, see Kai Lai Chung the
exitance of independent r.v.). The exitance of $X$ and $Y$ is
trivial since we can always define $X$ as identical mapping from
$(E,\mathcal{B}(E),\mu)$ to $(E,\mathcal{B}(E))$. Hence the coupling
of two probability measures does exit. \\\\
Remark 3: If $\mu$ and $\nu$ have the same distribution on $E$, then
there is a trivial coupling. Let $X$ be the r.v. with distribution
$\mu$, then the joint distribution of bivariate r.v. $(X,X)$ is the
trivial coupling of $\mu$ and $\nu$. In paper\cite{DWe05}, D. Weitz
always utilizes this coupling of two distributions with the same
distribution restricted on the same region. \\

A very important property of the coupling is that it can be used to
bound the distance in variation of two distributions. See the
following property.\\\\
\textbf{Property 2.1} Let $E$ be a countable state-space, and $\mu$
and $\nu$ be two probability measure on $E$. Let $X$ and $Y$ be two
random variables from the probability space on $(\Omega,
\mathcal{F}, P)$. Assume that $X$ and $Y$ have the distribution
$\mu$ and $\nu$. Then
\begin{displaymath}
d(\mu,\nu)\leq P(X\neq Y),
\end{displaymath}
where $d(\cdot,\cdot)$ denotes the distance in variation.\\\\
\textbf{Proof:} $\forall A\subset E$, there are
\begin{displaymath}
\begin{split}
P(X\neq Y)&\geq P(X\in A, Y\in A^{c})\\
&=P(X\in A) - P(X\in A, Y\in A)\\
&\geq P(X\in A) - P(Y\in A)\\
&=\mu(A)-\nu(A).
\end{split}
\end{displaymath}
Taking the supremum of $A$ in the previous inequality implies the
desired result.\\\\
Remark 1: If the state space $E$ is a metric space with metric $r$,
the following inequality follows quickly.
\begin{displaymath}
 P(X\neq Y)\leq\frac{
Exp(r(X,Y))}{\inf\limits_{i\neq j \in E}r(i,j)}.
\end{displaymath}
Combining this inequality and the one in the Property 2.1, we get
another bound of the distance in variation, which is the basis of
inequality (5) in the paper (Page 455).\\\\
Remark 2: The above property is one application of the coupling
method. Of course, it has  more wide applications. Just think about
the following example modified from the one given by Professor Zhan
Shi (I'll show you the proof in the class).\\\\
 \textbf{Problem 2.1} Let $X$ be a r.v. from $(\Omega, \mathcal{F},
 P)$ to $(R, \mathcal{B}(R))$. $f$ and $g$ are two monotone increase
 functions on $R$ a.s., then
\begin{displaymath}
E(f(X))E(g(X))\leq E(f(X)g(X)).
\end{displaymath}

I'm sorry I can't give the exact definition on ``path coupling",
however, I hope my illustration can help you grasp the essence of it.\\\\
Illustration: Suppose there're distributions $\mu_{1},
\mu_{2},\cdots, \mu_{n+1}$ on $E$. We already have the coupling of
$\mu_{j}$ and $\mu_{j+1}$ denoted by $L_{j}$, $j=1, 2, \cdots, n$.
How can we construct the coupling of $\mu_{1}$ and $\mu_{n+1}$ based
on this known coupling $L_{j}$, $j=1, 2, \cdots, n$? Precisely, we
need to construct a series of r.v. $X_{j}$, $j=1, 2, \cdots, n+1$
such that $L_{j}$ is the coupling of $X_{j}$ and $X_{j+1}$, $j=1, 2,
\cdots, n$. Then the distribution of $(X_{1},X_{n+1})$ is the
desired coupling of $\mu_{1}$ and $\mu_{n+1}$. The basic method is
to use conditional probability as working in the paper. Select an
element $x_{1}\in E$ according to the distribution $\mu_{1}$, then
select $x_{2}\in E$ according the distribution $L_{1}$ conditioned
on $x_{1}$. Now see what we have done. Since
$P(X_{1}=x_{1},X_{2}=x_{2})=P(X_{1}=x_{1})P(X_{2}=x_{2}|X_{1}=x_{1})=L_{1}(x_{1},x_{2})$.
taking the sum of $x_{1}$ over $E$, we know
$P(X_{2}=x_{2})=\mu_{2}(x_{2})$. Hence the above two steps have
construct two r.v. $X_{1}$ and $X_{2}$ with distribution $\mu_{1}$
and $\mu_{2}$ respectively and coupling $L_{1}$. Continue the above
settings, choose $x_{j}\in E$ according to $L_{j-1}$, $j=2,3,\cdots,
n+1$. Then we have constructed the r.v. $X_{j}$, $j=1, 2, \cdots,
n+1$ satisfying the previous requirement(one can check)since in one
stochastic experiment $X_{j}$, $j=1, 2, \cdots, n+1$ comes from the
same probability space. We can see this path coupling in some sense
decompose the comparison between two distributions(e.g. $\mu_{1}$
and $\mu_{n+1}$) into a series of sub-comparison(e.g.$\mu_{j}$ and
$\mu_{j+1}$, $j=1, 2, \cdots, n$), which may be easily calculated.
For example, if there is a metric $r$ on $E$, then
$Exp(r(X_{1},X_{n+1}))\leq
\sum\limits_{j=1}^{n}Exp(r(X_{j},X_{j+1}))$, which is the copy of
the inequality (6) in the paper (Page 456).
\section{details of formulas and some explanations}
Now I present the proofs or explanations of some of the formulas
with index which play an important role in understanding the paper.
The notations are the same as in the paper if there's no
specification. \\
\textbf{Erratum :} \\
1. Page 452 $\Theta_i \in B(x)$ should be $\Theta_{i \in B(x)}$ \\
2. Page 456 in the second paragraph \\$K_{S}(\eta^{(j-1)},\eta_{j})$ should be $K_{S}(\eta^{(j-1)},\eta^{(j)})$\\\\
\textbf{1. Formula (2)} Page 448\\\\
\textbf{Proof:} $\forall \tau = \sigma$ off $\Delta$
\begin{displaymath}
\begin{split}
\gamma_{\Lambda}^{\sigma}(\tau|\sigma_{\Delta^{c}})&=\frac{\gamma_{\Lambda}^{\sigma}(\tau,\sigma_{\Delta^{c}})}{\gamma_{\Lambda}^{\sigma}(\phi:\phi_{\Delta^{c}}=\sigma_{\Delta^{c}})}\\
&=\frac{\frac{1}{Z_{\Lambda}^{\sigma}}exp(-H_{\Lambda}(\tau))}{\sum\limits_{\phi:\phi_{\Delta^{c}}=\sigma_{\Delta^{c}}}\frac{1}{Z_{\Lambda}^{\sigma}}exp(-H_{\Lambda}(\phi))}\\
&=\frac{exp(-H_{\Lambda}(\tau))}{\sum\limits_{\phi:\phi_{\Delta^{c}}=\sigma_{\Delta^{c}}}exp(-H_{\Lambda}(\phi))}\\
&=\frac{exp(-H_{\Delta}(\tau))exp(-\bar{H}_{\Lambda/\Delta}(\tau))}{\sum\limits_{\phi:\phi_{\Delta^{c}}=\sigma_{\Delta^{c}}}exp(-H_{\Delta}(\phi))exp(-\bar{H}_{\Lambda/\Delta}(\phi))}\\
&=\frac{exp(-H_{\Delta}(\tau))exp(-\bar{H}_{\Lambda/\Delta}(\sigma))}{\sum\limits_{\phi:\phi_{\Delta^{c}}=\sigma_{\Delta^{c}}}exp(-H_{\Delta}(\phi))exp(-\bar{H}_{\Lambda/\Delta}(\sigma))}\\
&=\frac{exp(-H_{\Delta}(\tau))}{\sum\limits_{\phi:\phi_{\Delta^{c}}=\sigma_{\Delta^{c}}}exp(-H_{\Delta}(\phi))}\\
&=\frac{1}{Z_{\Delta}^{\sigma}}exp(-H_{\Delta}(\tau)),
\end{split}
\end{displaymath}
where
$\bar{H}_{\Lambda/\Delta}(\sigma):=\sum\limits_{x\in\Lambda/\Delta}U_{x}(\sigma_{x})+
\sum\limits_{\{x,y\}\in E: \{x,y\}\cap\Lambda\neq\emptyset,x\notin
\Delta, y\notin \Delta}U_{x,y}(\sigma_x,\sigma_y).$
\\\\
\textbf{2. Formula (3) and (4)} Page 449\\\\
\textbf{Proof:} For (3), you can understand $\mu_{1}(A)$,
$A\subset\mathcal{S}^{\Lambda}$ as $\mu_{1}(A\times\mathcal
{S}^{V/\Lambda})$.\\
For (4) in Weitz's proof, I explain ``the projection of $\mu$ on
$\mathcal{S}^{\Lambda}$ is a convex combination of the projections
of $\gamma_{\Psi}^{\sigma}$ on $\mathcal{S}^{\Lambda}$ as $\sigma$
varies." By the definition of Gibbs measure $\mu$, there are\\
\begin{displaymath}
\begin{split}
\mu(A\times\mathcal{S}^{V/\Lambda})&=\mu(A\times\mathcal{S}^{\Psi/\Lambda}\times\mathcal{S}^{V/\Psi})\\
&=\sum\limits_{\sigma}\mu(A\times\mathcal{S}^{\Psi/\Lambda}|\sigma_{\Psi^{c}})\mu(\phi:\phi_{\Psi^{c}}=\sigma_{\Psi^{c}})\\
&=\sum\limits_{\sigma}\gamma_{\Psi}^{\sigma}(A\times\mathcal{S}^{\Psi/\Lambda})\mu(\phi:\phi_{\Psi^{c}}=\sigma_{\Psi^{c}})
\end{split}
\end{displaymath}
Noting that the distance of any two points in a convex body is less
than the maximum over the distances of all pairs of vertices of it,
$\parallel\mu_1-\mu_2\parallel_{\Lambda}\leq
\sup\limits_{\tau,\sigma}\parallel\gamma_{\Psi_m}^{\tau}-\gamma_{\Psi_m}^{\sigma}\parallel_{\Lambda}$
follows quickly.\\\\
\textbf{3. Formula (5)} Page 455\\\\
\textbf{Proof:} See  Property 2.1 and its Remark 1, and note\\
\begin{displaymath}
\begin{split}
\rho_{\Lambda}(Q_m)=\sum\limits_{x\in\Lambda}\rho_x(Q_m)
&=\sum\limits_{x\in\Lambda}\sum\limits_{\eta_x\neq\xi_x}\rho_x(\eta_x,\xi_x)Q_m(\eta,\xi)\\
&=\sum\limits_{x\in\Lambda}\sum\limits_{\eta_{\Lambda}\neq\xi_{\Lambda}}\rho_x(\eta_{\Lambda},\xi_{\Lambda})Q_m(\eta,\xi)\\
&=\sum\limits_{\eta_{\Lambda}\neq\xi_{\Lambda}}\rho_{\Lambda}(\eta_{\Lambda},\xi_{\Lambda})Q_m(\eta,\xi)
\end{split}
\end{displaymath}
\\\\
\textbf{4. Formula (6)} \\\\
\textbf{Proof:} $E(\rho_{\Delta}(\sigma^{(0)},\sigma^{(n+1)}))\leq
E(\rho_{\Delta}(\sum\limits_{j=1}^{n+1}\sigma^{(j-1)},\sigma^{(j)}))$
and $E(\rho_{\Delta}(\sigma^{(n)},\sigma^{(n+1)}))=0$, then (6)
follows.\\\\
\textbf{5. Explanation of $F_{S}(Q)$ being a coupling} Page 456 in
the last paragraph. $F_{S}(Q)$ is a coupling of $\gamma_{\Psi}^{\sigma}$ and $\gamma_{\Psi}^{\tau}$. \\
\textbf{Proof:} $\forall \eta_1=\sigma, \eta_2 =\tau $ off $\Psi$.
then\\
\begin{displaymath}
\begin{split}
\sum\limits_{\eta_1}F_{S}(Q)(\eta_1,\eta_2)&=\sum\limits_{\eta_2}\sum\limits_{\eta,\xi}Q(\eta,\xi)K_{S}(\eta,\xi)(\eta_1,\eta_2)\\
&=\sum\limits_{\eta,\xi}\sum\limits_{\eta_2}Q(\eta,\xi)K_{S}(\eta,\xi)(\eta_1,\eta_2)\\
&=\sum\limits_{\eta,\xi}Q(\eta,\xi)\kappa_{S}^{\eta}(\eta_1)\\
&=\sum\limits_{\eta}\kappa_{S}^{\eta}(\eta_1)\sum\limits_{\xi}Q(\eta,\xi)\\
&=\sum\limits_{\eta}\kappa_{S}^{\eta}(\eta_1)\gamma_{\Psi}^{\sigma}(\eta)\\
&=w_S^{-1}\sum\limits_{\eta}\sum\limits_{i\in
S}w_i\kappa_{i}^{\eta}(\eta_1)\gamma_{\Psi}^{\sigma}(\eta)\\
&=w_S^{-1}\sum\limits_{i\in
S}\sum\limits_{\eta}w_i\gamma_{\Psi}^{\sigma}(\eta_1|\eta_{\Theta_i^c})\gamma_{\Psi}^{\sigma}(\eta)\\
&=w_S^{-1}\sum\limits_{i\in
S}\sum\limits_{\phi}w_i\gamma_{\Psi}^{\sigma}(\eta_1|\eta_{\Theta_i^c})\gamma_{\Psi}^{\sigma}(\phi:\phi_{\Theta_i^c}=\eta_{\Theta_i^c})\\
&=w_S^{-1}\sum\limits_{i\in S}w_i\gamma_{\Psi}^{\sigma}(\eta_1)\\
&=\gamma_{\Psi}^{\sigma}(\eta_1).
\end{split}
\end{displaymath}
From this, we also know $F_{S}^{t}$ is a coupling of
$\gamma_{\Psi}^{\sigma}$ and $\gamma_{\Psi}^{\tau}$.
\\
\textbf{5. Formula (12)}\\\\
\textbf{Proof:} Noting that
\begin{displaymath}
\begin{split}
\rho_{\Delta}(K_i(\eta^{(j-1)},\eta^{(j)}))&\leq
\rho_{\Delta/\Theta_i}(K_i(\eta^{(j-1)},\eta^{(j)})) +
\rho_{\Delta\cap\Theta_i}(K_i(\eta^{(j-1)},\eta^{(j)})) \\
&=\rho_{\Delta/\Theta_i}(\eta^{(j-1)},\eta^{(j)})+\rho_{\Delta\cap\Theta_i}(K_i(\eta^{(j-1)},\eta^{(j)}))\\
&=\rho_{z_j}(\eta^{(j-1)},\eta^{(j)})1_{z_j\in\Delta/\Theta_i}+\rho_{\Delta\cap\Theta_i}(K_i(\eta^{(j-1)},\eta^{(j)}))
\end{split}
\end{displaymath}
\begin{center} {and}\end{center}
\begin{displaymath}
\begin{split}
\rho_{\Delta}(K_S(\eta^{(j-1)},\eta^{(j)}))&=w_S^{-1}\sum\limits_{i\in
S}w_i\rho_{\Delta}(K_i(\eta^{(j-1)},\eta^{(j)}))\\
&=w_S^{-1}(\sum\limits_{i\in B(z_j)}+\sum\limits_{i\in
S/B(z_j)})w_i\rho_{\Delta}(K_i(\eta^{(j-1)},\eta^{(j)}))
\end{split}
\end{displaymath}
Using Weitz's explanations, Formula(12) follows.


\begin{references} 


\bibitem{DWe05}
{Dror Weitz}.
\newblock{Combinatorial Criteria for Uniqueness of Gibbs Measures},
\newblock {\emph{Random Structures and Algorithms}.}  (2005), 445-475.


\end{references}

\end{document}